\documentclass[journal, doublecolumn, 10pt]{IEEEtran}

\usepackage{times}
\usepackage[latin1]{inputenc}
\usepackage[T1]{fontenc}
\usepackage[english]{babel}
\usepackage{balance}

\usepackage{amsmath, amsfonts, amssymb, array,  booktabs}
\usepackage{subeqnarray}
\usepackage{algorithmic, algorithm}
\usepackage{amstext}

\usepackage{graphicx}
\usepackage{caption}

\usepackage{psfrag}
\usepackage{epsf,epsfig}
\usepackage{subfigure}

\usepackage{multirow}
\usepackage{url}
\usepackage{color}
\usepackage[normalem]{ulem}
\usepackage{enumerate}

\begin{document}
%

\title{Learning Radio Resource Management in RANs:\\ Framework, Opportunities and Challenges}


\author{\IEEEauthorblockN{\normalsize Francesco Davide Calabrese\IEEEauthorrefmark{1}, Li Wang\IEEEauthorrefmark{1}, Euhanna Ghadimi\IEEEauthorrefmark{1}, Gunnar Peters\IEEEauthorrefmark{1},  Lajos Hanzo\IEEEauthorrefmark{2}, Pablo Soldati\IEEEauthorrefmark{1}}
	\\
\IEEEauthorblockA{
\IEEEauthorrefmark{1}Huawei Technologies Sweden AB, Wireless Network Algorithms Lab, Kista, Sweden\\
}
\IEEEauthorblockA{\IEEEauthorrefmark{2}
University of Southampton, School of ECS, Wireless Research Group, Southampton, UK
}
}
\maketitle
\pagestyle{empty}
\thispagestyle{empty}

\begin{abstract}
In the fifth generation (5G) of mobile broadband systems, Radio Resources Management (RRM) will reach unprecedented levels of complexity. To cope with the ever more sophisticated RRM functionalities and with the growing variety of scenarios, while carrying out the prompt decisions required in 5G, this manuscript presents a lean 5G RRM architecture that capitalizes on recent advances in the field of machine learning in combination with the large amount of data readily available in the network~from measurements and system observations.
The architecture relies on a single general-purpose learning framework conceived for RRM directly using the data gathered in the network. The complexity~of~RRM~is~shifted~to~the design of the framework, whilst the RRM algorithms derived from this framework are executed in a computationally efficient distributed manner at the radio access nodes. The potential of this approach is verified in a pair of pertinent scenarios and future directions on applications of machine learning to RRM are discussed.
\end{abstract}


\section{Introduction}\label{sec:1}

Radio Resource Management (RRM) in Radio Access Networks (RANs) is a large-scale control problem encompassing numerous network functionalities operating at different time-scales ranging from sub-millisecond to seconds. The architecture governing RRM in today's RANs is the result of incremental engineering, with new RRM functionalities constantly being added to follow the system evolution. While this approach has stimulated rapid development of the 3GPP LTE system, ten years down the road of system evolution lead to an increasingly fragmented RRM architecture founded on an ever growing number of parameters.

The complexity of an RRM task depends on the dimensionality of the problem at hand and the available execution time. With the fifth generation (5G) of mobile broadband systems, which shall integrate new technology components (e.g. massive MIMO, mm-Wave communication, network slicing, vehicular networks, more and broader frequency bands, etc.) with larger optimization domains and tighter latency requirements, RRM is expected to reach unprecedented complexity~\cite{ABC:14}-\nocite{BHL:14}\cite{AIS:14}. Thus, optimizing such large-scale RRM problems~with traditional rule-based algorithms is particularly challenging.

RANs are data-rich environments, where data~is continuously gathered in the form of radio measurements or other system observations by thousands of user devices and network entities. Nonetheless, RRM nowadays derives little insight from such data. On the one hand, data is utilized as input to run rule-based algorithms then swiftly discarded either upon aging (e.g., after few milliseconds) or when a user moves across radio cells. On the other hand, the decisions taken using such data are circumscribed in time and space. As a result, RANs currently treat data as a short-lived and localized commodity.

The rapid advances in the Machine Learning (ML) field, however, combined with the recent technology leap in hardware specialized to handle large data sets, present an opportunity to give data a more central role in wireless networking.

The contributions of this manuscript are twofold: A logical architecture to enable an efficient implementation of ML based solutions in RANs and a general purpose learning framework capable of autonomously generating, directly from data, algorithms specialized for the RRM functionality at hand.
Thus, the learning framework treats data available in the RAN as a growing source of information from which RRM algorithms are derived (i.e., learned) and refined over time.
We consider a framework broadly based on Reinforcement Learning (RL), a branch of ML suitable for solving  control problems such as the ones arising in RRM.
Although basic RL formulations have been successfully applied to specific RRM functionalities (see, e.g.,~\cite{GaG:10}\nocite{BeN:10}-\cite{MFO:13}), a general-purpose application of RL to RRM poses additional challenges beyond the scope of these studies.
Starting from such challenges, we discuss architectural and algorithmic approaches that can overcome the challenges and enable general purpose learning for RRM in RANs.

{Although the ideas proposed in this manuscript are described in the context of radio cellular networks, with 5G being the natural application scenario, the concepts extend to any RAN technology where one can conceive the idea of a central unit gathering data from the edge nodes. For instance, a proprietary Wi-Fi network deployed by a network operator, could collect data from the access network and learn to adjust several parameters of the MAC or PHY payer, such as the contention window, the thresholds for user association, transmission power, etc.}

\section{A brief overview of learning methods}\label{sec:2}
Machine learning deals with the task of inferring a function from a set of noisy data, known as the training set, generated by an unknown true function. The ML branches of interest, here, are supervised learning~\cite{DeepLearnBook:16} and reinforcement learning~\cite{SuB:98}.

Supervised learning infers a function from a set of data pairs comprising an input and a desired output, provided by a supervisor. Artificial Neural Networks (ANNs) are a class of function approximators which can be tailored to the task at hand through proper adjustment of their weights. Training an ANN consists in gradually modifying its weights in the direction minimizing an error function between the function represented by the ANN and the actual noisy data samples produced by the original true function. The application of the gradient method, specialized to ANNs, is known as back-propagation~\cite{DeepLearnBook:16}.

Reinforcement learning, on the other hand, deals with how a software agent learns to behave in an environment to achieve a given objective, e.g., maximizing a form of reward. Thus, it is particularly suitable for control problems, such as those arising in RRM. Hereafter we consider a model-free setup, where the problem is described exclusively in terms of three components: \emph{state}, \emph{action} and \emph{reward}.

The state $s$ is a tuple of values, known as \emph{features}, that characterizes the environment for the agent in a way that is relevant to the problem at hand. The action $a$ represents the change that the agent applies to the environment. The reward $r$ is a multi-objective scalar function which numerically expresses the agent's purpose. The interaction, over time, of the agent with the environment is captured by a set of tuples $(s_t,a_t,r_{t+1},s_{t+1})$, where $t$ is a discrete time counter, describing a state transition as a consequence of applying actions to the environment and receiving rewards.

The objective of RL is to extract, from a set of transitions, a policy $\pi$ that, given a state, returns the action to take in order to maximize the long-term cumulative reward. An RL algorithm thus maps the rewards to actions, possibly taken far back in time. This notion is known as \emph{credit-assignment}~\cite{SuB:98}.

An RL algorithm shall rapidly bring an agent from a tabula rasa state, where it does not know how to act, to a condition where it acts as close to optimality as possible. Making as few mistakes as possible in the path to a quasi-optimal behavior is known as \emph{regret minimization}, a notion closely related to the topic of trading off \emph{exploration} of the environment (to sample unseen parts of the state-action space at the cost of not choosing the best known action for a state) with \emph{exploitation} of the knowledge accumulated so far (to maximize the reward at the cost of not trying a new potentially better action). This gradual transition from a pure exploration strategy to an exploitation strategy can be implemented using a variety of techniques, e.g. $\epsilon$-greedy algorithm~\cite{SuB:98}.

\section{RL for RRM: challenges and opportunities } \label{sec:3}

Successful application of RL to the RRM in modern RANs is bound to addressing a number of challenges arising from the intricate and highly dynamic nature of such environments.

\subsection{Challenges}\label{sec:challenge}

The first challenge is the large dimensionality of RRM problems in RANs. The variety of conditions in the network, paired with the number of configurable parameters, results in an extremely large state-action space.
In 5G systems, for instance, the cardinality of the RRM decision domains is affected by massive number of connected devices, larger number of operating bands with wider bandwidths, flexible sub-frame lengths (from $1$ ms down to $125$ $\mu$s) and sub-carrier spacing, etc.~\cite{ABC:14}-\nocite{BHL:14}\cite{AIS:14}. Moreover, the stringent requirements for latency-critical applications reduce the execution time of RRM functionalities, e.g. for scheduling resources, to less than $100$ $\mu$s, thus increasing the hardness of the task even further.

Another crucial challenge for applying RL to RANs is the partial observability of the network state available to the agent, as provided by local measurements taken by users in a cell or by the access nodes controlled by the baseband board unit (BBU) where an agent resides.
New computationally demanding radio access technologies (RATs), such as Massive MIMO and mm-Wave, will reduce the cells controllable by a BBU, thus the agent's observability.

Additionally, learning a global policy in large-scale multi-agent environments like RANs poses strong challenges in terms of coordination among agents.
Finally, a challenge of practical nature pertains how to control the exploration process so as to prevent a prolonged degradation of the system performance and, therefore, user experience.

\subsection{Opportunities}

More powerful computer hardware technology, more efficient data storage technology, and more advanced machine learning tools present an opportunity to radically rethink the design of RRM algorithms.
The leap in Graphical Processing Units (GPUs) and multi-core Central Processing Units (CPUs) technology made massive parallel computing widely available at relatively low cost. This gives the opportunity to consider the abundance of data continuously gathered in radio networks as the source of RAN's intelligence from which RRM algorithms can be derived and progressively updated, {as opposite to today's approach of generating data to run RRM functionalities and discarding it soon thereafter.
Data can therefore be reused both spatially (between cells) and in time. For instance, data associated to users that left the network is still valuable as it encodes system experience which can be reused at a later time or in other parts of the network.}

Another opportunity is the drive towards extensive radio measurements, e.g. originating from user centric uplink beaconing sensed with large antenna arrays at the network side, enabling more exact positioning and radio finger printing. Radio Environment Maps (REM) can offer convenient ways of representing such data. Additionally, the networks gather data related to both user mobility and to traffic patterns, or more generally as to how/when/what the user acts in the network, at much higher~rate~than~ever~before.


\section{Learning architecture for RRM} \label{sec:4}

\begin{figure}[t!]
\centering{
\includegraphics[width=1\linewidth]{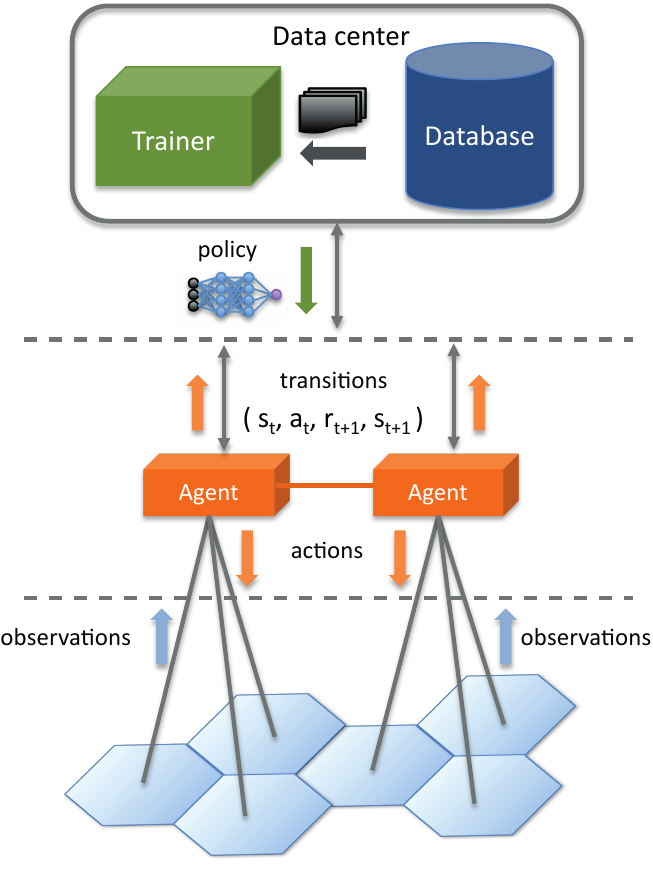}
}
\caption{Learning architecture to enable centralized training and the distributed execution at the agents in RANs.}
\label{fig:1}
\end{figure}

To embrace these opportunities and enable fundamental learning techniques for RRM in RANs, we propose the architecture in Figure~\ref{fig:1}, comprising two logical entities, namely the agent and the trainer, overlaying existing or upcoming RATs.

Mapping Figure~\ref{fig:1} to ML parlance, the RAN represents the environment. The state, in its entirety, can be represented by a set of features characterizing the agent in relation to the network obtained from system observations in the form of, e.g., radio measurements and protocol measurements, such as signal-to-noise ratio, interference measurements, data traffic, resource block utilization, spectral efficiency, as well as network characteristics such as the type and capabilities of terminals, traffic patterns, the type and number of cells, and system key performance indicators (KPIs) (e.g., cell coverage, cell capacity, packet delay) etc. In practice, each RRM functionality is associated with a more compact state containing a small set of relevant features. The actions specific to an RRM functionality (e.g., power control) are represented by parameters adjustments (e.g., power-up, power-down, power-hold). The reward may represent a function (typically non-linear) of conventional KPIs or system requirements used in wireless networking. For instance, the harmonic mean of the user rates can be used to balance between coverage and capacity requirements (cf.~\cite{GCP:16}).

\subsection{The trainer-agent split}
The trainer and the agent are the key components of the architecture. The trainer applies a single learning algorithm to generate control algorithms (policies, in RL parlance) specialized for different RRM functionalities. Agents execute policies issued by the trainer in a distributed fashion to control and interact with the network. While traditionally co-located, we argue that the trainer, where policies are derived offline, and the agent, where policies are executed in real-time, should reside in different network entities, as shown in Figure~\ref{fig:1}.

The rationale for the split is two-fold: firstly, central training enables efficient transfer learning (cf. Sec.~\ref{transfer_learning}), a crucial tool to overcome many of the challenges identified in Section~\ref{sec:challenge}. This enables each access node, e.g. Gigabit-NodeB (gNB) in 5G, to benefit from the experience gathered by other access nodes; secondly, the amount of computational resources needed for training (from hours to days) and execution (fraction of a ms) might be entirely different and require entirely different hardware dimensioning. As such, carrying out training and inference at the same place would require replicating expensive hardware types (e.g., GPUs for the trainer and CPUs for the agent) across the network.

\subsection{Communication aspects}
The trainer communicates policy updates to the agent. The agent communicates data transitions to the trainer. While~a~policy update may consist of a new set of parameters (e.g., ANN weights, cf. Sec.~\ref{sec:trainer}), the data transitions consists of a set of tuples describing the network state, an action taken by the agent, and the reward and new state observed upon taking the action.

Both policy and data transitions are exchanged over the network backhaul without affecting the air interface resources. A new policy could be issued at regular time intervals (hourly/daily/weekly) or based on some dynamic criteria, e.g. the amount of new data that needs to be collected before a new training is initiated or the time required by the training algorithm itself to produce a new policy. The data transitions are created on a faster time-scale, (e.g., tens of ms to minutes) but can be exchanged in large batches on a best effort basis. Neither communication directions are therefore delay sensitive.

The learning architecture further suggests an interface between agents to enable inter-agent coordination, e.g., the \emph{Xn} interface in a 5G system. The type of information exchanged over this interface can be task-specific, such as observations of inter-cell interference to each agent to mitigate interference. We will return to these aspects in Sec.~\ref{sec:6bobservability}.

\subsection{Implementations aspects}
A natural implementation is to consider one (or a few) central trainer in the network (e.g., a small-scale server farm) with the agents residing within radio access nodes, e.g. at BBU. The RRM algorithms are thus generated centrally, while their execution is distributed across individual radio access nodes.

In alternative, the agent could reside in the BBU pool of a Cloud-RAN (C-RAN) architecture. In this case, the C-RAN receives the RRM algorithms from the trainer, executes the algorithms and forwards the decisions (actions in RL parlance) to the controlled radio heads.
While one may be tempted to co-locate the trainer and agent at the C-RAN BBU pool, this would require to also store large amount of data directly in the C-RAN hardware, resulting in higher costs and losing the advantages of the trainer-agent split.

\section{Learning algorithmic framework} \label{sec:5}

\begin{figure*}[t!]
\centering{
\includegraphics[width=0.8\linewidth]{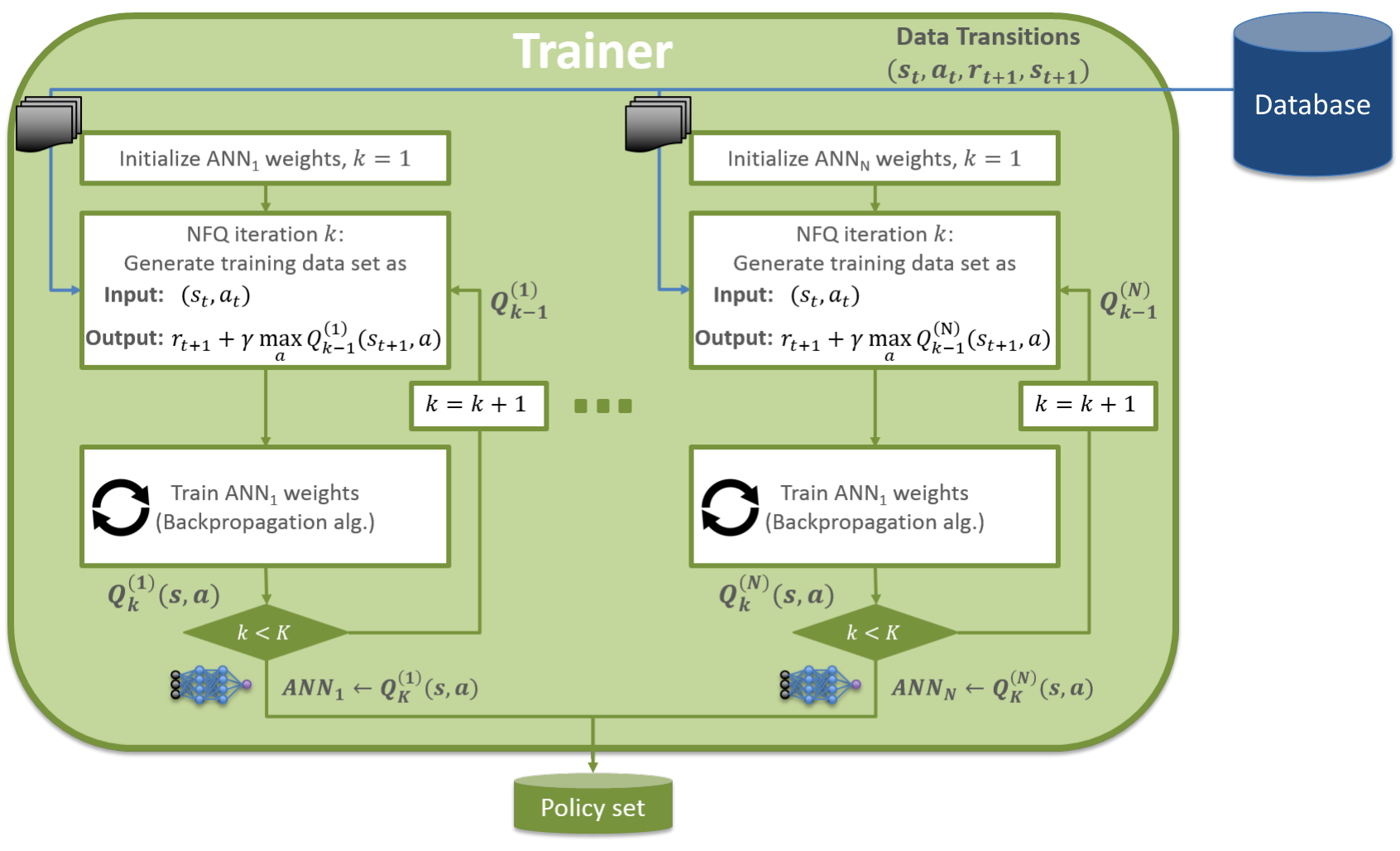}
}
\caption{Illustration of functional blocks of a general purpose RRM trainer design with policy generation procedure based on Q-learning with NFQ iteration for an ensemble of $N$ ANNs.}
\label{fig:2}
\end{figure*}
To provide a concrete example of how the challenges of learning RRM algorithms in a radio environment could be addressed, we propose a general learning framework consisting of three major components: Neural-Fitted Q-Iteration (NFQ), ensemble learning and transfer learning.
While alternative and more advanced learning frameworks may be designed already today, the intention of this work is to exemplify how one such framework would be deployed in the proposed architecture to produce control policies specialized to individual RRM functionalities. In other words, the architecture in Sec.~\ref{sec:4} is based on principles that are independent from the algorithmic framework adopted which is, therefore, replaceable.

\subsection{RRM trainer design}\label{sec:trainer}

\begin{figure*}[t!]
\centering{
\includegraphics[width=1\linewidth]{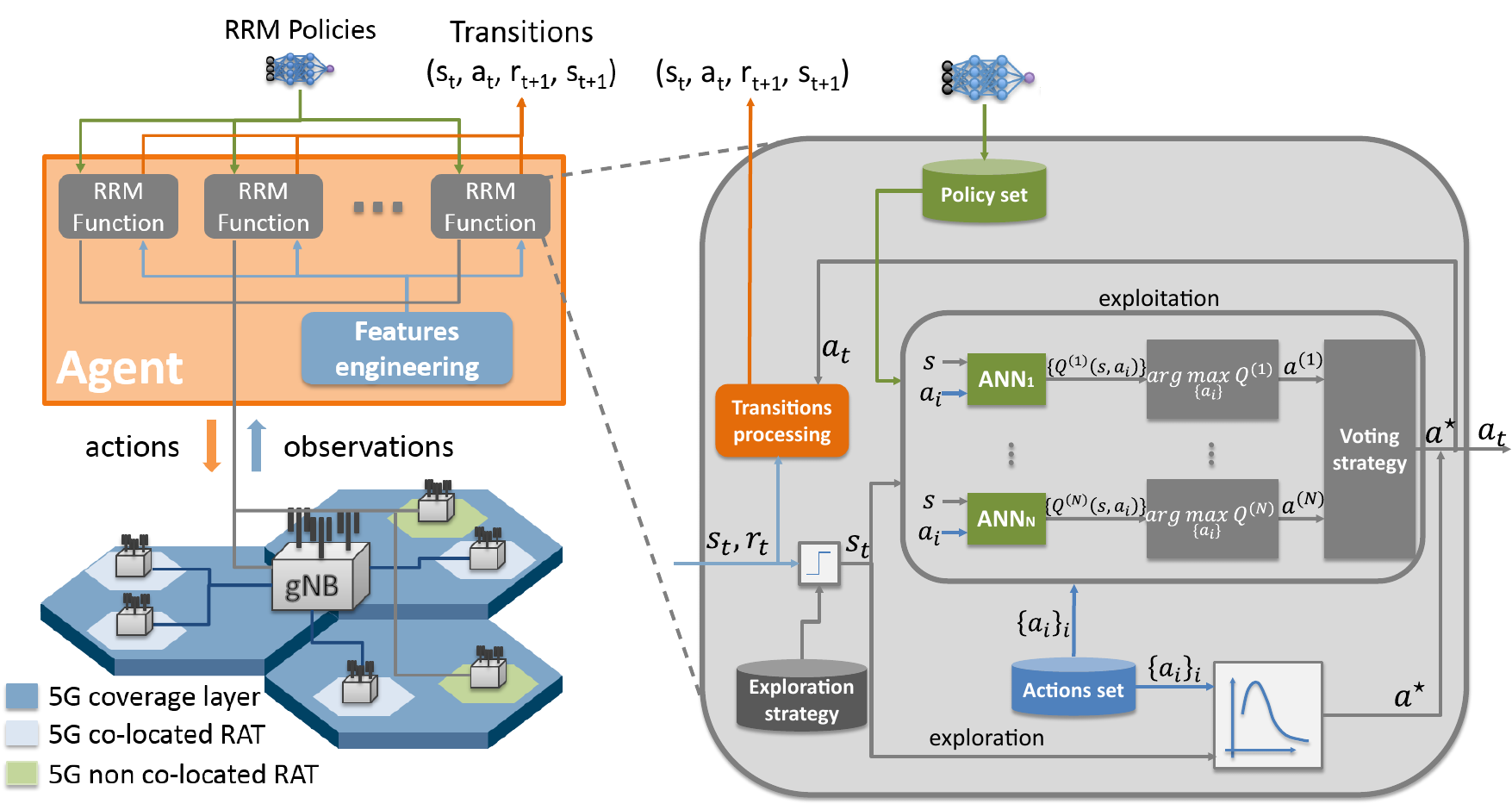}
}
\caption{Illustration of a multi-purpose RRM agent design with decision process of an agent based on an exploration-exploitation strategy and an ensemble of ANNs.}
\label{fig:3}
\end{figure*}

A common reinforcement learning approach for control problems is to learn an action-value function, e.g. the Q-function, to estimate the long-term reward of a state-action pair. Thus, for a given a policy $\pi$, the Q-function seen by an agent in a state $s$ at time $t$, taking the action $a$, and thereafter following policy $\pi$, is defined as the expected sum of the discounted future rewards
\begin{equation}
\nonumber
\begin{aligned}
Q_{\pi} (s, a ) = \mathbb{E}_\pi \Big[\sum_{l=t}^\infty {\gamma^{l-t} R_{l+1}}\mid S_t=s, A_t=a\Big],
\end{aligned}
\end{equation}
where $R$ denotes a reward function, dependent on both state and action, and $\gamma\in[0,1)$ is a discounting factor used to adjust the preference for immediate reward\footnote{We follow the notation in~\cite{SuB:98} and remark the difference between random variables and their instantiations with capital letters and lower letters, respectively. Thus, the state, action, and reward are denoted by $S$, $A$, and $R$, respectively, their possible values by $s$, $a$, and $r$. We also use subscript $t$ to indicate a time step $t$ and a superscript $'$ to indicate a successor state or action.}.
Connecting the reward to the KPIs of the system allows to optimize the system performance over a shorter ($\gamma\rightarrow 0$) or longer ($\gamma\rightarrow 1$) horizon.

Maximizing the long-term reward is equivalent to searching a policy that maximizes the Q-function. When the control problem can be modeled as Markov decision process, the optimal Q-function satisfies the Bellman optimality equation~\cite{SuB:98}
\begin{equation}
\nonumber
\begin{aligned}
Q^\star(s, a) =  \mathbb{E}_{S'\sim\mathcal{P}} [R  +\gamma  \;\underset{a'}{\max} \;Q^\star(S', A')\mid S=s,A=a],
\end{aligned}
\end{equation}
where $\mathcal{P}$ denotes the state transition probability distribution of the environment. RL algorithms essentially sample the environment and learn the Q-function by using the Bellman equation as an iterative update, e.g.,
\begin{align*}
Q_{k+1}(s, a) =\mathbb{E} [r  +\gamma  \;\underset{a'}{\max} \;Q_k(s',a')\mid S=s, A=a],
\end{align*}
where $k$ denotes an iteration index and $\mathbb{E}$ an empirical mean over samples of experience, with $r\sim R$.

The simplest approach to Q-learning is to store and update $Q$-values for individual state-action pairs in a table~\cite{SuB:98}. While tabular-based Q-learning has also been suggested for wireless applications (see, e.g.,~\cite{GaG:10}\nocite{BeN:10}-\cite{MFO:13}), this approach is unsuitable for RRM due to the large dimensionality of the state-action space.

\subsubsection{Neural-Fitted Q-iteration}
To cope with the large dimensionality of RRM problems, we suggest to apply Q-learning via functional approximation of the Q-function~\cite{SuB:98}. Using supervised learning techniques (cf. Sec.~\ref{sec:2}), this approach directly learns a function $Q(s, a, w)$ parameterized by a set of parameters $w$ that are optimized by minimizing a proper cost function, i.e., a measure of mismatch between the current Q-value $Q(s,a)$ and the target Q-value $r  +\gamma\, \underset{a'}{\max} \,Q(s',a')$.

Although different functional approximators could be used within the learning framework, we suggest ANNs due to their excellent generalization capabilities and the existence of computationally efficient training algorithms~\cite{DeepLearnBook:16}.
To train the ANN, we suggest the Neural-Fitted Q-iteration algorithm~\cite{GLR:11}. NFQ is an iterative approach which, given a set of transitions $\{(s_t,a_t,r_{t+1},s_{t+1})\}_t$, constructs a training set consisting of state-action pairs $(s_t ,a_t)$ as input and the Q-target $[r_{t+1}  +\gamma \; \underset{a_{t+1}}{\max} \;Q(s_{t+1},a_{t+1})]$ as desired output to train the ANN.

Figure~\ref{fig:2} shows the NFQ policy generation procedure. Starting with a random initialization, an ANN $n$ is trained over $K$ iterations to output $Q^{(n)}=Q_K^{(n)}$. At each iteration $k$, the Q-function $Q_k^{(n)}$ is improved as follows: an outer loop generates the training data based on the Q-function $Q_{k-1}^{(n)}$ from the previous iteration and the set of transitions; an inner loop trains the weights $w^{(n)}$ of the ANN to fit the newly generated input-output pairs to return $Q_k^{(n)}$.

\subsubsection{Ensemble learning}
To improve generalization, we suggest to adopt \emph{ensemble learning} where $N$ ANNs with distinct structures and configurations (e.g., number of layers, neurons per layer, weights initialization, etc.) are independently trained to learn alternative Q-functions from the same set of data (see Figure~\ref{fig:2}). {As long as each ANN in the ensemble chooses the correct action more than $50\%$ of the times and the ANNs are uncorrelated with each other, aggregating multiple ANNs at the agent via a properly designed voting strategy (e.g., majority voting) results in a superior policy.}

\subsubsection{Transfer learning}\label{transfer_learning}
Wireless networks are always-running critical systems. It is therefore crucial to learn (i.e., generalize) in a sample efficient way, e.g. with fewer samples possible.
The trainer-agent split enables sample efficiency by crowd-sourcing data across the network and sharing the resulting experience across nodes via transfer learning~\cite{LeaningToLearn:12}.

Two basic forms of transfer learning are \emph{parameter transfer} and \emph{instance transfer}. The former consists in sharing the same policy (e.g., using the same ANN weights) across different agents, thus significantly reducing the number of free parameters learned from a data set. The latter consists in sharing experience (i.e., data transitions) among agents in the network.
A variant of instance transfer consists in initially providing the trainer with artificial data samples generated from a network simulator which allows to jump-start the learning.

\subsection{RRM agent design}

The agent interacts with the underlying cellular network by executing the RRM algorithm (an ensemble of ANNs) for one or more controlled radio cells, cf.~Figure~\ref{fig:3}.

For each RRM functionality, the agent follows an exploration strategy to take either an explorative action or an  exploitative action selected by running majority voting on the ensemble of ANNs, cf.~Figure~\ref{fig:3}.
%
The $\epsilon$-greedy algorithm~\cite{SuB:98} is a simple, yet effective, method wherein the agent
randomly explores the state-action space by taking actions randomly from a given distribution.
Expert human knowledge can also be used to confine the exploration within more relevant regions of the state-action space so as to improve sample efficiency.

Upon taking an action, the agent processes the system observation into a data transition of the form $(s_t,a_t,r_{t+1},s_{t+1})$ and transfers it to the trainer.

\section{How the challenges are addressed}\label{sec:6}
The trainer-agent split combined with the suggested learning framework helps to address several of the challenges of learning RRM algorithms in a radio environment.

\subsection{Large state space dimensionality}

The proposed learning framework addresses the large state dimensionality with functional approximation of the Q-function, which allows to generalize from seen to unseen states.

A related problem is to achieve a faithful representation of the state via the input data to the functional approximator with low communication overhead
One approach is to let the trainer learn a representation of the state by extracting state features directly from high-dimensional raw observations based, for instance, on deep learning methods~\cite{DeepLearnBook:16}. This has the advantage of producing an accurate state estimation while keeping the logic for learning the model at the data
center. However, the communication overhead to stream raw observations from the agent to the trainer can quickly become prohibitive. We leave a more thorough investigation of this approach to a future study.

The approach here presented, instead, known as \emph{feature engineering}, consists in identifying and crafting a large set of informative features which the agent streams up to the trainer. The trainer can further process these features to achieve a compact yet representative state. This approach typically leads to a less accurate state recovery but also faster convergence of the policy and lower bandwidth requirements on the agent-trainer interface.

\subsection{Complexity and fast execution}

The trainer-agent split enables the proper dimensioning of the hardware depending on the task at hand, i.e. training vs execution. While the computational and storage requirements are entirely shifted to few central training nodes, the intelligence, encoded by the policy, is retained at the distributed agents.

Additionally, using ANNs in the learning framework allows to capitalize the massively parallel computation models of modern hardware (e.g., GPUs for general-purpose computing) to accelerate both the training and the execution of complex RRM tasks at the agent, potentially within the stringent time requirements of 5G.

\subsection{Partial observability and multi-agent coordination}\label{sec:6bobservability}
The availability of richer observations made available by new radio access technologies, such as Massive MIMO, per se improves the agent's understanding of the local environment. Additionally, by expanding the state representation with observations collected at different times one may capture temporal correlation in changes affecting the state of the environment.

Observability can further be improved beyond the agent's surrounding by inter-agent communication and coordination, for instance by integrating an agent's state with some knowledge of the state of neighbouring agents. This allows to take actions that maximize a collective reward, such as those necessary for interference coordination, instead of individual rewards. This approach was used in the study cases of Sec.~\ref{sec:evaluation}.

\begin{figure*}[t!]
\centering{
\includegraphics[width=1\linewidth]{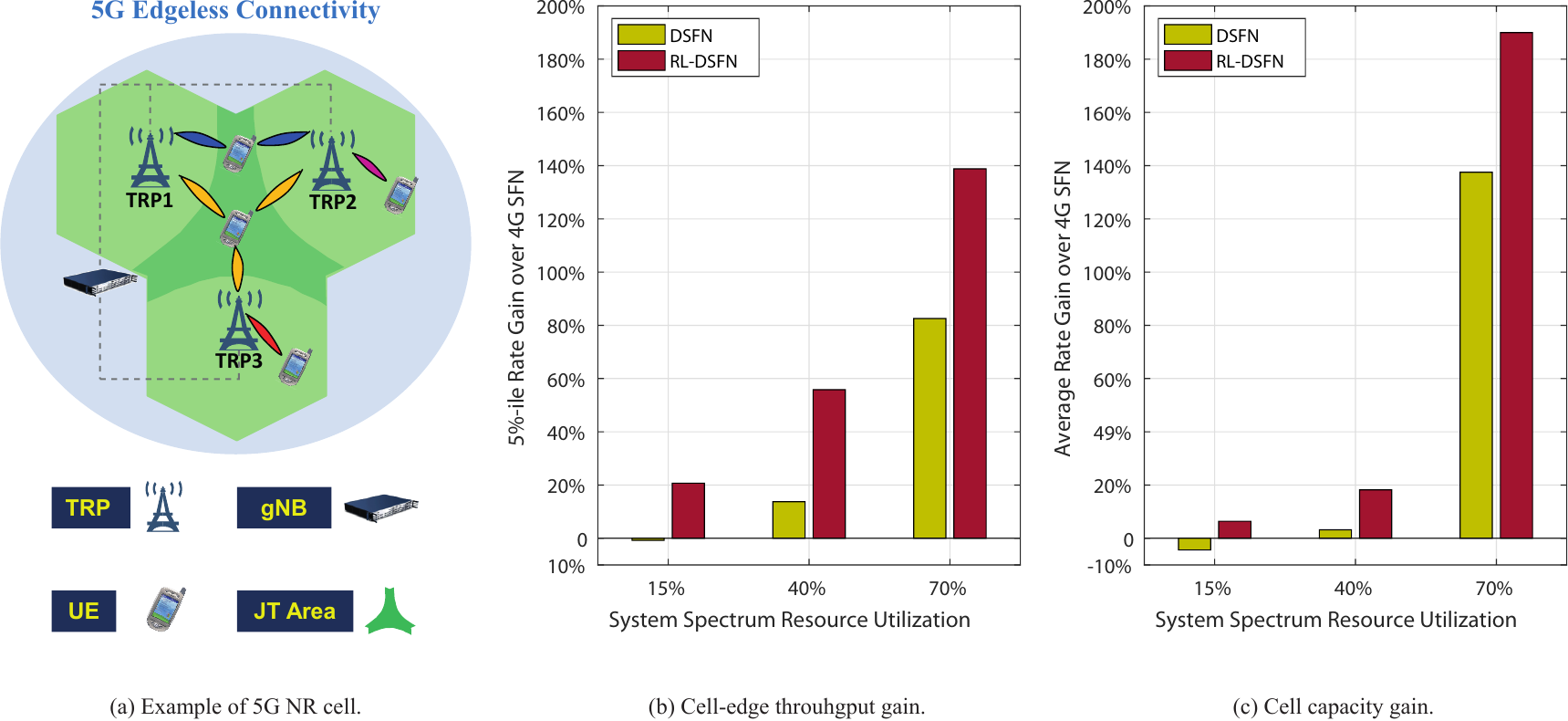}
}
\caption{User centric TRP selection and joint transmission for edgeless connectivity in 5G NR.}
\label{fig:4}
\end{figure*}

\subsection{Network heterogeneity}
{To handle the heterogeneity of modern radio cellular systems in terms of coexisting radio access technologies and frequency bands (i.e. different types of radio cells), it is important to properly categorise the network observations. This enables the trainer to learn different policies for a common task (e.g. power control) to different type of cells.}
{Categorising the observations per type of cell, bandwidth, RAT, etc, may be useful, for instance, to
learn coarse estimates of the channel spatial parameters at high frequencies, where the cost for observations is high in terms of radio radio resources utilization, by using out-of-band observations at lower frequency bands, such  as communication systems operating at sub-6GHz bands, sensors and positioning information, which are available with lower radio resource utilization~\cite{Gonzalez:17}}.

\subsection{Scalability}

In addition to scaling well with the large dimensionality and complexity of the RRM problems, the trainer-agent split brings the added benefit of scalable communication overhead. On one hand, enabling transfer learning allows reusing experience from different agents to quickly collect rich data sets with fewer samples from individual agents. This can also reduce the need for extensive exploration. Transfer learning can also be used to warm-start newly deployed agents with a good policy.
On the other hand, an agent with a well performing policy can significantly reduce the frequency of data exchange towards the trainer until a significant change in performance occurs. Conversely, the trainer can reduce policy update frequency, with the twofold benefit of reducing communication overhead and training cost.

{Finally, engineering features at the agent, as opposed to learning features at the trainer from raw data\iffalse with a deep learning framework\fi, allows to transfer a more compressed version of the observation and better scale with the bandwidth limitation of the network backhaul.}

\section{Evaluation study cases}\label{sec:evaluation}

\begin{figure*}[t]
  \centering
  \subfigure[CDF of mean user rate.]
  {\includegraphics[width=0.32\hsize]{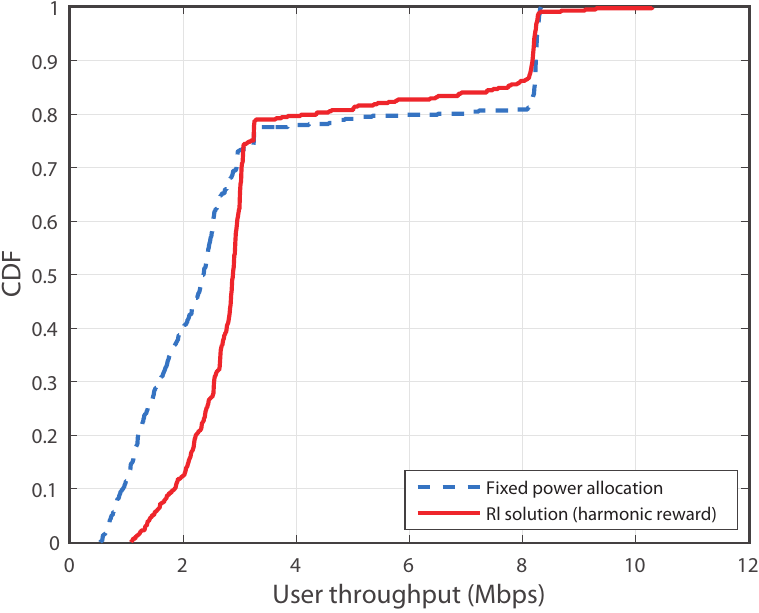}
  \label{fig:cdf_bursty}}
  \subfigure[CDF of the network throughput.]
  {\includegraphics[width=0.32\hsize]{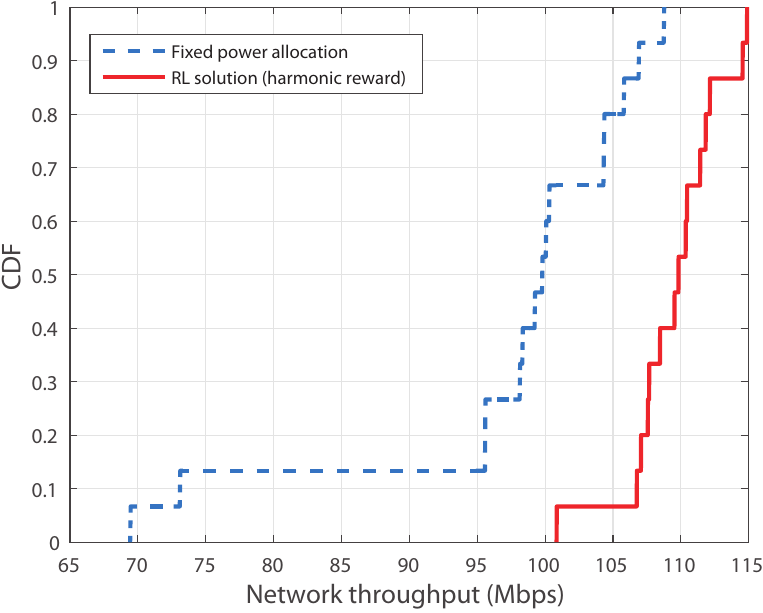}
  \label{fig:power_bursty}}
  \subfigure[CDF of the network throughput normalized by the transmission power (log-scale)]
  {\includegraphics[width=0.32\hsize]{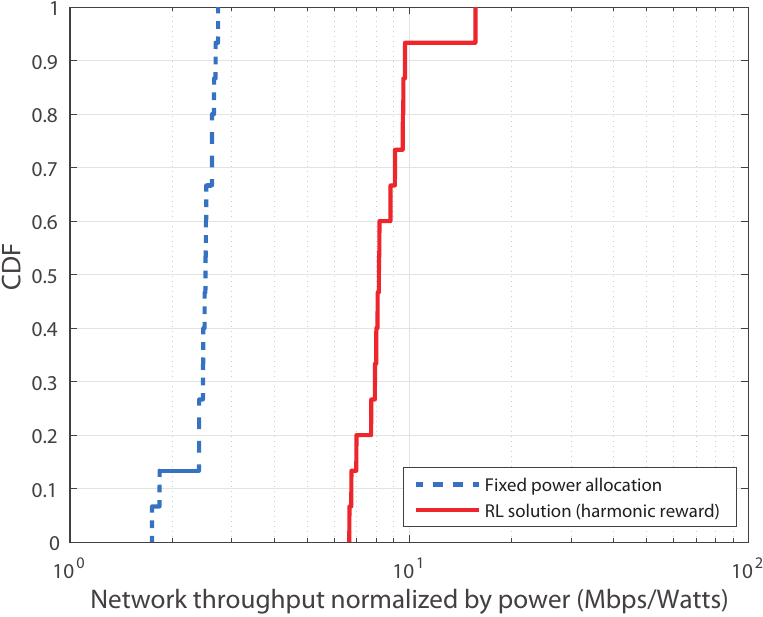}
  \label{fig:thru_bursty}}
  \caption{System level evaluation of RL-based power control and rate adaptation with bursty traffic.} \label{fig:MBBM1_bursty_traffic}
\end{figure*}

We illustrate the learning framework using two RRM problems, one from 5G and one relevant to any interference-limited cellular system.
We evaluate them in a sub-6GHz event-driven system simulator and compare the results to state-of-the-art RRM functionalities.

\subsection{Edgeless connectivity}

3GPP defines a 5G New Radio (NR) cell as a gNB connected to multiple transmission/reception points (TRPs), each providing NR synchronization signals carrying~the~same~cell ID. While conceptually a NR cell may resemble Single-Frequency Network (SFN) operation in 4G, its aim is to enable advanced user-centric features and edgeless connectivity by means of joint transmission/reception, coordinated scheduling,~etc.

Removing cell edges requires to address two problems: first, associating users to multiple TRPs for coordinate multi-point (CoMP) operation - a NP-hard task even for single TRP case~\cite{Ye:UserAssociationLoadBalancingHetnet};  second, dynamically forming joint transmission areas between TRPs by optimizing multiple SIR-thresholds per TRP to adjust to changes in traffic load and user mobility.

We solve this problem with end-to-end learning exploiting the general-purpose RRM learning framework of Sec.~\ref{sec:5}. The system is characterized by a set of features constructed from pre-processed measurements, such as the first and second order statistics of the resource utilization per TRP as well as the traffic load's geographical distribution, etc. The system reward is represented by the harmonic mean of the user rates. The expected reward is optimized by adjusting a SIR-threshold (the action) which determines the TRPs participating in the JT.

Figure~\ref{fig:4} shows that the proposed algorithm, in red, learns to adapt the SIR-thresholds to the time-varying load distributions, and significantly improves both the coverage and capacity compared to basic SFN operation (user associated to all nodes) and an enhanced SFN with dynamic user associated to one or more nodes depending on a geometry-like factor (DSFN). The simulation results are obtained based on real network traffic in Singapore generated by both small and big packet users.

\subsection{Distributed downlink power control}

Careful load balancing and efficient inter-cell interference mitigation will be crucial in ultra-dense small cell deployments~\cite{ABC:14}. We therefore studied the joint downlink power control and rate adaptation by using our RL framework adapted to a multi-agent setting, where each agent controls a cell.

To compress the dimensionality of the learning space, the state of the network is partially observed via a few carefully selected features such as the cells power budget, average SINR, and sum user rates.
While the features are extracted from local measurements within a cell, a global network-wide reward is reconstructed by means of inter-agent exchange of the local cell rewards, which encourages the emergence of cooperative behaviors among agents.

Figure~\ref{fig:MBBM1_bursty_traffic} shows the simulation results for a multi-cell scenario supporting bursty traffic. Compared to a baseline with fixed power budget, end-to-end learning achieves significant throughput gains at both the $5\%$-tile and the median users (cf. Figure~\ref{fig:cdf_bursty}), thereby improving fairness across cells and the overall network throughput (cf. Figure~\ref{fig:power_bursty}).
Even more significant are the gains in power reduction which, for each cell, range from 3 dB to 6 dB. Therefore, when the network throughput is normalized by the transmission power, the gap in performance becomes even larger, as shown in Figure~\ref{fig:thru_bursty}. We refer to~\cite{GCP:16} for a more detailed description.

\section{Final remarks and future directions}\label{sec:conclusion}

The abundance of data and significant improvements in capabilities of modern hardware and ML algorithms make the time ripe for introducing a fundamentally different RRM architecture, where individual rule-based RRM algorithms are replaced by a general purpose learning framework capable of autonomously generating, directly from data, complex algorithms specialized for the RRM functionality at hand.

The advantages of this approach are multi-fold. Firstly, the experience (i.e. data) gathered by an access node can be reused to improve the behavior of other nodes.
Secondly, improving the single learning framework leads to an improvement across all RRM tasks, resulting in a compounded RRM performance gain.
Thirdly, a new node installed in the operator's network will be promptly equipped with a near optimal policy by benefiting from the  experience gathered by existing nodes.
Moreover, changes in the non-stationary wireless environment are automatically taken into account by continuous learning at the network side. These benefits naturally result in significant Capital Expenditure (CAPEX) and Operating Expense (OPEX) reductions, while enhancing the system performance.

The proposed learning NFQ-iteration combined with transfer learning and ensemble learning is an example of framework capable of improving the performance against the state-of-the-art  for a variety of RRM functionalities.
Nonetheless, the fast pace of advances in the ML field regularly brings about more powerful techniques that can enrich or replace parts of the framework without affecting the overall architecture, which is intended to be future-proof.
One such extension, given the recent successes of deep learning, is to directly learn features from raw measurements. While this approach requires significantly more data samples, naturally available in modern RANs, than the current feature-engineered solutions,  the resultant features would certainly be more accurate.
Another interesting and more involved extension is to jointly solve multiple RRM tasks with the aid of a single learned algorithm by extending the number of actions the agent has access to.

\balance
\bibliography{Reference}
\normalfont

\end{document}